\documentclass[aip,jcp,reprint,twocolumn,floatfix,eqsecnum]{revtex4-1}

\usepackage{amsmath}
\usepackage{amssymb}
\usepackage{graphicx}
\usepackage{mathrsfs}

\begin{document}

\title{Breathing dynamics based parameter sensitivity analysis of hetero-polymeric DNA}

\author{Srijeeta Talukder}
\affiliation{Department of Chemistry, University of Calcutta, 92 A P C Road, Kolkata 700 009, India}

\author{Shrabani Sen}
\affiliation{Department of Chemistry, University of Calcutta, 92 A P C Road, Kolkata 700 009, India}

\author{Prantik Chakraborti}
\affiliation{Department of Chemistry, Bose Institute, 93/1 A P C Road, Kolkata 700 009, India}

\author{Ralf Metzler}
\email{rmetzler@uni-potsdam.de}
\affiliation{Institute for Physics \& Astronomy, University of Potsdam,
D-14476 Potsdam-Golm, Germany}
\affiliation{Physics Department, Tampere University of Technology,
FI-33101 Tampere, Finland}

\author{Suman K Banik}
\email{skbanik@jcbose.ac.in}
\affiliation{Department of Chemistry, Bose Institute, 93/1 A P C Road, Kolkata 700 009, India}

\author{Pinaki Chaudhury}
\email{pinakc@rediffmail.com: Corresponding author}
\affiliation{Department of Chemistry, University of Calcutta, 92 A P C Road, Kolkata 700 009, India}

\date{\today}

\begin{abstract}
We study the parameter sensitivity of hetero-polymeric DNA within the purview of DNA 
breathing dynamics. The degree of correlation between the mean bubble size and the
model parameters are estimated for this purpose for three different DNA 
sequences. The analysis leads us to a better understanding of the sequence
dependent nature of the breathing dynamics of hetero-polymeric DNA. 
Out of the fourteen model parameters for DNA stability in the statistical
Poland-Scheraga approach, the hydrogen bond interaction $\epsilon_{hb}(\mathtt{AT})$ for
an $\mathtt{AT}$ base pair and the ring factor $\xi$ turn out to be the most sensitive
parameters. In addition, the stacking interaction $\epsilon_{st}(\mathtt{TA}-\mathtt{TA})$ for an $\mathtt{TA}-\mathtt{TA}$ nearest neighbor pair of base-pairs is
found to be the most sensitive one among all stacking interactions. 
Moreover, we also establish that the nature of stacking interaction has a 
deciding effect on the DNA breathing dynamics, not the number of times a particular 
stacking interaction appears in a sequence.
We show that the sensitivity analysis can be used as an effective measure to guide
a stochastic optimization technique to find the kinetic rate constants related to
the dynamics as opposed to the case where 
the rate constants are measured using the conventional unbiased way of optimization.
\end{abstract}


\maketitle

\section{Introduction}

Hydrogen bonding between the complementary base pairs ($\mathtt{AT}$ and $\mathtt{GC}$)
is the origin for the Watson Crick double helical DNA structure \cite{watson}. The secondary interaction via the 
nearest neighboring stacking interaction also has a major contribution towards the DNA structure. 
The base pair stacking compensates the repulsive electrostatic force of  phosphate groups of the two 
complementary bases which come closer due to hydrogen bonding, henceforth giving stability to 
the helical conformation. Although this double helix is the most stable form of DNA, it is not a static one 
\cite{frank,wartell,grosberg,poland1,poland}. The hydrogen bonds can intermittently open up and rejoin, 
even at room temperature and normal salt concentration, without damaging the core of the 
nucleotide. This transient denatured zone in a DNA polymer is commonly known as a bubble. 
As the total energy needed to open up a base pair depends on the nature of that base 
pair (hydrogen bond interaction) as well as its neighborhood (stacking interaction), the 
probability of bubble formation becomes a function of the DNA sequence, i.e it is connected 
to the stability profile of a genome. It is also important to mention that in most natural DNA, 
the opening probability is much higher due to torsional stress \cite{bauer,jeon}.
The breathing dynamics of bubble plays a crucial role in the functioning of DNA 
\cite{pollock,pant,ambj1,sokolov,yeramian,choi,nowak,zoli,phelps}. 
Fundamental biological processes 
like replication and transcription largely rely on the local denaturation. 
A recent study on the interaction between the nucleoid-associated protein Fis and DNA
in \textit{E. coli} suggests that Fis-DNA interaction is controlled by DNA breathing dynamics
and can be regulated experimentally via different nucleotide modifications \cite{nowak}.
The physical properties of the DNA direct the biological functioning of the living system. DNA breathing is 
thus a good problem to study, both with respect to its physical and biological perspective.

Many experimental techniques, such as, circular dichroism \cite{cantor},
UV spectroscopy \cite{cantor}, calor\-imetry \cite{senior}, fluorescence
resonant energy transfer (FRET) measurements \cite{gelfand} can map out
the melting profile of DNA. Using single molecule florescence correlation
spectroscopy, breathing dynamics has been monitored and multistep relaxation
kinetics with characteristic time scale has been accounted from the study
\cite{altan,kathy}.  Experimental studies have also been performed to explore
correlations between the dynamics of hetero-polymeric DNA with the biological
activities of nucleic acid enzymes \cite{phelps}. Theoretical models of the
dynamics have also been established based on the DNA free energy landscape
\cite{hanke,bicout,bar,novotny,fogedby}.  Another way of studying breathing
dynamics is by carrying out a stochastic simulation \cite{ambj2,ambj3}
using the Gillespie Algorithm \cite{gill1,gill2} .

Sequence sensitivity is one of the pivotal motivations for studying DNA
breathing dynamics. From the time series data of the dynamics, information
about DNA sequence and its stability parameters can be estimated.
Single DNA manipulation techniques can produce stability parameters and
can account for the strong dependence on salt concentration of the breathing
dynamics \cite{huguet}.  In one of our previous communications we showed that
the conjunction of breathing dynamics of hetero-polymeric DNA with
one of the stochastic optimization technique, namely Simulated Annealing,
can provide data regarding the stability parameters, as well as the
activation energy and critical exponent with good accuracy \cite{talukder1}.
However, the dependence of the breathing dynamics on the individual parameters
had not been discussed in earlier work. We here ask whether can we quantify the
relative influence of the system parameters on the breathing dynamics? To answer
this question we quantify the sensitivity of the stability parameters as well as
the activation energy and the critical exponent with respect to the breathing
dynamics of a hetero-polymeric DNA using the approach of sensitivity analysis
(SENSA).

SENSA is used to understand the relative importance of input 
parameters with respect to the system output. Generally SENSA is of two types, local and 
global \cite{saltelli,marino}. Local SENSA is based on gradient calculation and can account for the local effect of that particular input parameter on the output, for which the calculation is performed. 
It generally fails to provide accurate results when all the input parameters come 
into consideration simultaneously. The global SENSA follows a statistical formulation. The 
fundamental theory regarding the global strategy is how the variance of the output is guided by the perturbation on individual input parameters. Global SENSA is thus a sampling based technique. There are various ways to calculate the global SENSA and the success of these techniques are system specific. One of the two most popular method is the measure of correlations between the input and output parameters which is essentially used for systems whose output varies monotonically with the input. But for those systems which do not follow a monotonic trend, the decomposition of the
variance represents the best choice for determining sensitivity index. The most reliable variance based method is eFAST proposed by Saltelli et. al. \cite{saltelli1}. 
eFast is actually based on the Fourier Amplitude Sensitivity Test (FAST), developed by Cukier et.al \cite{cukier,cukier1} and Schaibly et.al \cite{schaibly}.  In the variance based sensitivity test the partitioning of  variance (of output) is done for determining what fraction of the variance in the output occurs due to the variation of each input parameters. This is known as the partial variance of output. The sensitivity of a particular input parameter is estimated using the ratio of the partial variance of the output (for that particular input) to the total output variance.

SENSA has a wide range of application in different fields like economics \cite{pannell}, environmental science \cite{hornberger}, systems biology \cite{zi1}, or chemical kinetics \cite{saltelli}. Biologists use SENSA to understand the robustness of the model output with respect to the variation in model inputs. This study also helps to analyze the dynamical behavior of the biological model.
In chemistry, there is also a long history of applying SENSA in chemical kinetics.
The SENSA of rate constants on the 
reaction kinetics is an important way to understand a kinetic scheme \cite{tohsato,talukder2}. It also has profound implications on parameter optimization \cite{talukder2}.

We use the correlation coefficient as a measure of the the index of sensitivity
for the parameters associated with the breathing dynamics as the breathing
dynamics is monotonically related to these parameters. In this communication
the Pearson correlation coefficient (CC), the rank correlation coefficient
(RCC), and the partial rank correlation coefficient (PRCC) \cite{book,hora}
are calculated for three different DNA sequences and their performance compared.
We also discuss the relevance of the SENSA on the parameter optimization and how
these results can dramatically influence an optimization process.
Specifically,
we study how the DNA breathing parameters, as described by the Poland-Scheraga
model, respond when subjected to a SENSA test. In other words, the gradation
of these model parameters on the basis of sensitivity will give us a
picture as to which of the model parameters play a more important role
in controlling the breathing dynamics. We also pursue the
important question whether the sensitivity order is dependent
on the nature of the sequence. In an earlier work,
stochastic optimization (Simulated Annealing) was demonstrated to extract
reliably the interaction energies in DNA breathing \cite{talukder1}. In the
present work we show that taking the sensitivity of individual parameters into
consideration, an optimization procedure becomes more efficient in the
determination of interaction energies of DNA breathing data. The Genetic Algorithm,
which has a completely different philosophy of operation as that of
Simulated Annealing, is used as the stochastic optimizer in this
work. One of the reason of using Genetic Algorithm over Simulated Annealing
is that the Genetic Algorithm, because of the relatively large search
space it can sample and exploit, requires a smaller number of optimization steps
to converge than that of Simulated Annealing \cite{talukder2}.

\section{Breathing Dynamics in DNA}

The stability of the double helical DNA hetero-polymer can be explained
by considering the two types of Watson-Crick hydrogen bond interaction
between the complementary bases $\mathtt{A}$ and $\mathtt{T}$, and $\mathtt{G}$ 
and $\mathtt{C}$ as well as the ten types
of stacking interactions between the nearest neighbor base pairs. In numbers,
the net free energy released due to opening of a base pair,
whose nearest neighbor base pair to one side is already denatured, is quite low,
due to the fact that enthalpy cost and entropy gain almost cancel.
Thus the free energy involved to break the strongest
interaction, a $\mathtt{GC}$ base pair stacked with a $\mathtt{CG}$ downstream of
the DNA sequence, is around $\sim 3.9$ $k_BT$ (at 37$^0$), whereas the denaturation
of an $\mathtt{AT}$ base pair with a downstream $\mathtt{TA}$ is marginally unstable with free energy change $\sim 0.1$
$k_BT$ (at 37$^0$) \cite{krueger}. The unzipping of DNA double helix is an
entropy driven process as it is basically a transformation from an ordered
to disordered conformation. The high binding enthalpy is compensated by
the entropy gain. But for a bubble initiation the activation energy is very
high, of the order of 7-12 $k_BT$ (for weakest and strongest respectively) as
breaking of two stacking interactions along with the disruption of a hydrogen
bond are concerned. Thus it is justified to assume that bubble events are
rare and two bubbles are well separated below the melting temperature.

A bubble formation event may be denoted by the position of the left zipper
fork ($x_L$) and the size of the bubble ($m$) in terms of the right zipper
fork $x_R=x_L+m+1$. One can visualize the breathing dynamics as a random
walk of a bubble on a triangular lattice of $x_L$ and $m$ with forbidden
horizontal transition. The Master equation depicting this process
\begin{equation}
\label{meq}
\frac{\partial P(x_L,m,t)}{\partial t}=\mathbb{W}P(x_L,m,t),
\end{equation}
where $P(x_L,m,t)$ is the probability of the occurrence of a bubble of size $m$ 
at the left zipper fork $x_L$ at a time $t$ and $ \mathbb{W} $ matrix include all the allowed 
transition rates from the state $(x_L,m)$ in the triangular lattice. The transfer rates are defined 
in terms of the Boltzmann factor of hydrogen bonding interaction and stacking interaction
\begin{equation}
\label{swth}
u_{hb}(x)=\exp\left(\frac{\epsilon_{\mathrm{hb}}(x)}{k_BT}\right),
u_{st}(x)=\exp\left(\frac{\epsilon_{\mathrm{st}}(x)}{k_BT}\right).
\end{equation}
In Eq.~(\ref{swth}), $ u_{hb}(x) $ is the Boltzmann factor for hydrogen bond at base 
pair position $x$ and $ u_{st}(x) $ is the Boltzmann factor for nearest neighbor stacking interaction between the 
base pairs at $x-1$ and $x$, respectively.
At $t\rightarrow\infty$, $P(x_L,m,t)$ from Eq.~(\ref{meq}) equilibrates to a probability distribution 
obtained from the statistical mechanical Poland-Scheraga model \cite{poland1,poland} of the DNA double 
helix. The equilibrium probability of a bubble of size $m$ and left zipper fork position $x_L$ can be 
written as
\begin{equation}
\label{eqd}
P_{\mathrm{eq}}(x_L,m)=\frac{\mathbb{Z}(x_L,m)}{\mathbb{Z}(0)+\sum_{m=1}^M
\sum_{x_L=0}^{M-m}\mathbb{Z}(x_L,m)},
\end{equation}
where $\mathbb{Z}(x_L,m)$ is the bubble partition function 
\begin{equation}
\label{pf}
\mathbb{Z}(x_L,m)=\frac{\xi'}{(1+m)^c}\prod_{x=x_L+1}^{x_L+m}u_{\mathrm{hb}}(x)
\prod_{x=x_L+1}^{x_L+m+1}u_{\mathrm{st}}(x).
\end{equation}
Eq.~(\ref{pf}) is for bubble size $m>0 $. If $m=0$, $\mathbb{Z}(0)=1$. Moreover,
$\xi'= 2^c\xi$, where $ \xi $ is the ring factor which contributes to the cooperativity factor. The cooperativity factor is the free energy cost for bubble activation. 
The ring factor is the key element for the formation of a small constrained loop in DNA double helix \cite{krueger}.
Finally, $c$ is the critical exponent and is related to the entropy factor during bubble formation
\cite{kaiser2014}.
The term $(1+m)^{-c}$ accounts for the loss of entropy during the formation 
of a polymer loop. From the probability expression one may write the equilibrium mean bubble 
size $\langle m \rangle$, for a sequence of base pair of length $M$ as
\begin{equation}
\label{mb}
\langle m \rangle = \frac{\sum_{m=1}^{M}m\sum_{x_L=0}^{M-m}\mathbb{Z}(x_L,m)}{\mathbb{Z}(0)+\sum_{m=1}^M\sum_{x_L=0}^{M-m}\mathbb{Z}(x_L,m)} .
\end{equation}

\section{Sensitivity Analysis of the DNA Stability Parameters}

Sensitivity analysis is an efficient tool to understand the degree of
susceptibility of the system with respect to different input parameters.
The biological functions of DNA are actually affected by the bubble dynamics as
intermittent bubble opening of base pair helps in binding of RNA polymerase, or
single-stranded DNA binding proteins, etc. Thus it is relevant to study the
effect of the stability parameters on the breathing dynamics. To this end
we use SENSA as a tool to measure the different weights of
the DNA breathing model parameters (hydrogen bond interaction $\epsilon_{hb}$
and stacking interaction $\epsilon_{st}$) on the process. As mentioned in the
previous section, the equilibrium probability distribution for bubble formation
also involves the ring factor factor $\xi$ and the critical exponent factor
$c$, which should have significant effect on bubble formation. This leads us
to consider all fourteen model parameters in the SENSA, namely, two hydrogen bond
interactions, ten nearest neighbor stacking interactions, as well as the ring
factor and critical exponent.

In order to quantify the effect of all fourteen parameters
on the breathing dynamics of hetero-polymeric DNA, we consider the mean
bubble size $\langle m \rangle $ as the measure for the breathing
dynamics, which can be quantitatively determined via experiment. The
breathing dynamics of hetero-polymeric DNA is sensitive to the DNA sequence
\cite{ambj2}, which leads us to consider three different DNA sequences,
the promoter sequence of the T7 phage \cite{ambj2}, as well as the L42B12
\cite{zeng} and AdMLP \cite{choi} sequences, for the present study to get a complete picture of sensitivity
of the breathing parameters. The promoter sequence of T7 phage is used as
the first sequence and is represented as
\begin{eqnarray}
\label{T7}
\texttt{5'-aTGACCAGTTGAAGGACTGGAAGTAATACGACTC} \nonumber \\
\texttt{AGTATAGGGACAATGCTTAAGGTCGCTCTCTAGGAg -3'} . \nonumber \\
\end{eqnarray}

\begin{figure}[!t]
\includegraphics[width=0.75\columnwidth,angle=0,clip=]{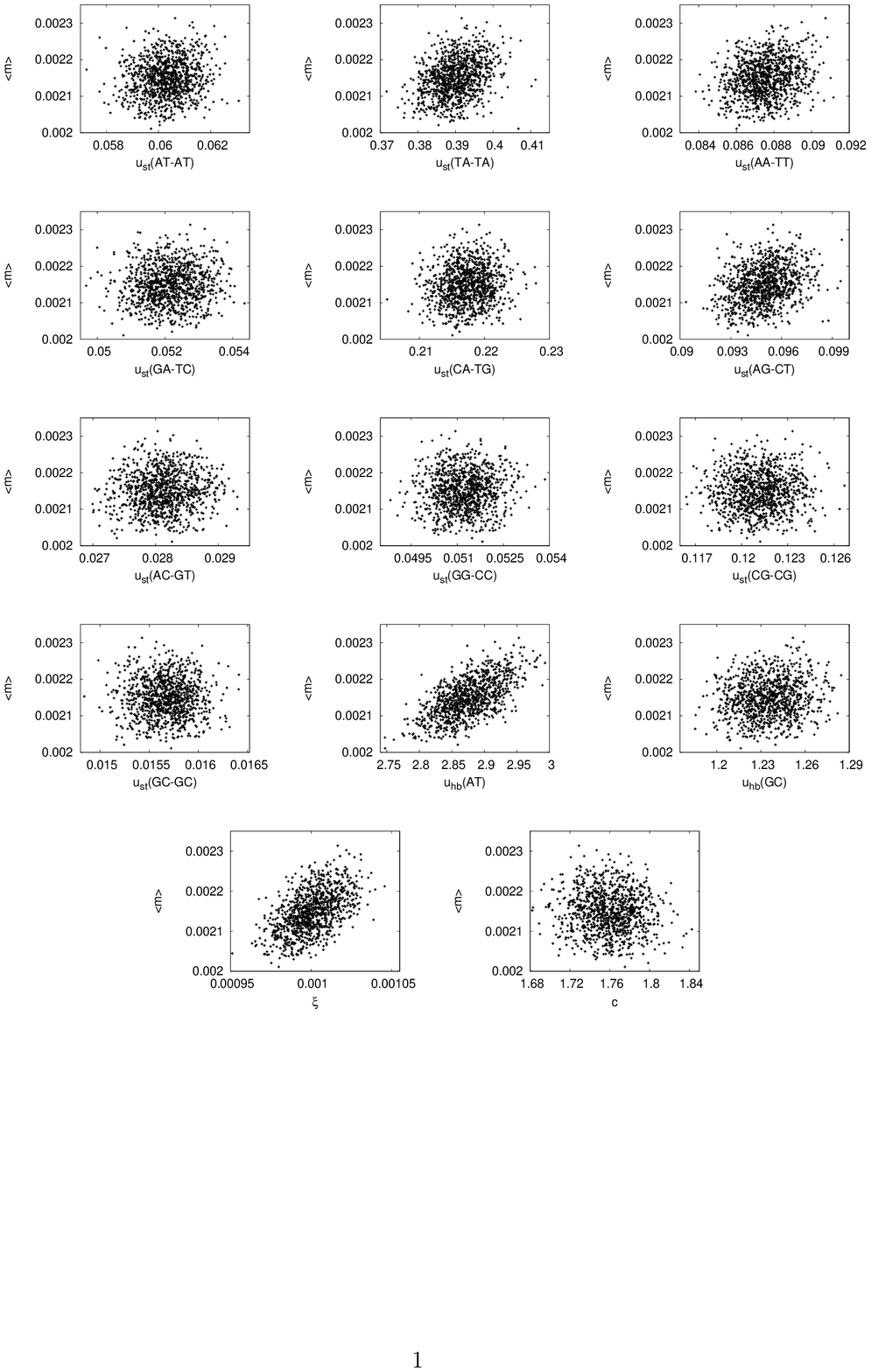}
\caption{Scatter plots of the mean bubble size versus the parameters associated
with the breathing dynamics of hetero-polymeric DNA, for the promoter sequence
of the T7 phage.}
\label{fig1}
\end{figure}

Firstly, to start the SENSA test, we generate scatter plots
of the mean bubble size vs all the parameters for the promoter sequence of T7 
phage, which gives a first hand qualitative picture of sensitivity.
To generate a scatter plot for a particular input parameter against an 
output, all the input parameters of the system under consideration are 
perturbed simultaneously and the output is calculated using the set of 
perturbed parameters.
If the output thus generated using the perturbed set of parameters is
plotted (typically known as input vs output scatter plot) against the
values of one specific parameter and falls in 
a narrow region around a virtual straight line, the corresponding input 
parameter seems to be more important with respect to that output parameter 
and is considered to have higher degree of sensitivity. Points dispersed in a circular 
region in the scatter plot denotes less or no correlation between the 
input and the output.
Fig.\ref{fig1} represents the set of the scatter plots generated using
the above mentioned procedure for the promoter sequence of T7 phage.
The input data has been picked up by a random process with the mean 
situated at the reported value (of that model parameter) \cite{krueger,richard} and the width 
of the perturbation being $ \pm $5$ \% $ of the reported value.
For most of the input parameters, plots appear as a dispersed set of points. 
Only in the scatter plots for the ring factor ($\xi$) and $\epsilon_{hb}(\mathtt{AT})$, the 
distribution of points falls in a narrow strip (or follow a definite direction) and
hence could be considered to be more sensitive parameters related to DNA
breathing dynamics.


\begin{figure}[!b]
\includegraphics[width=1.0\columnwidth,angle=0,clip=]{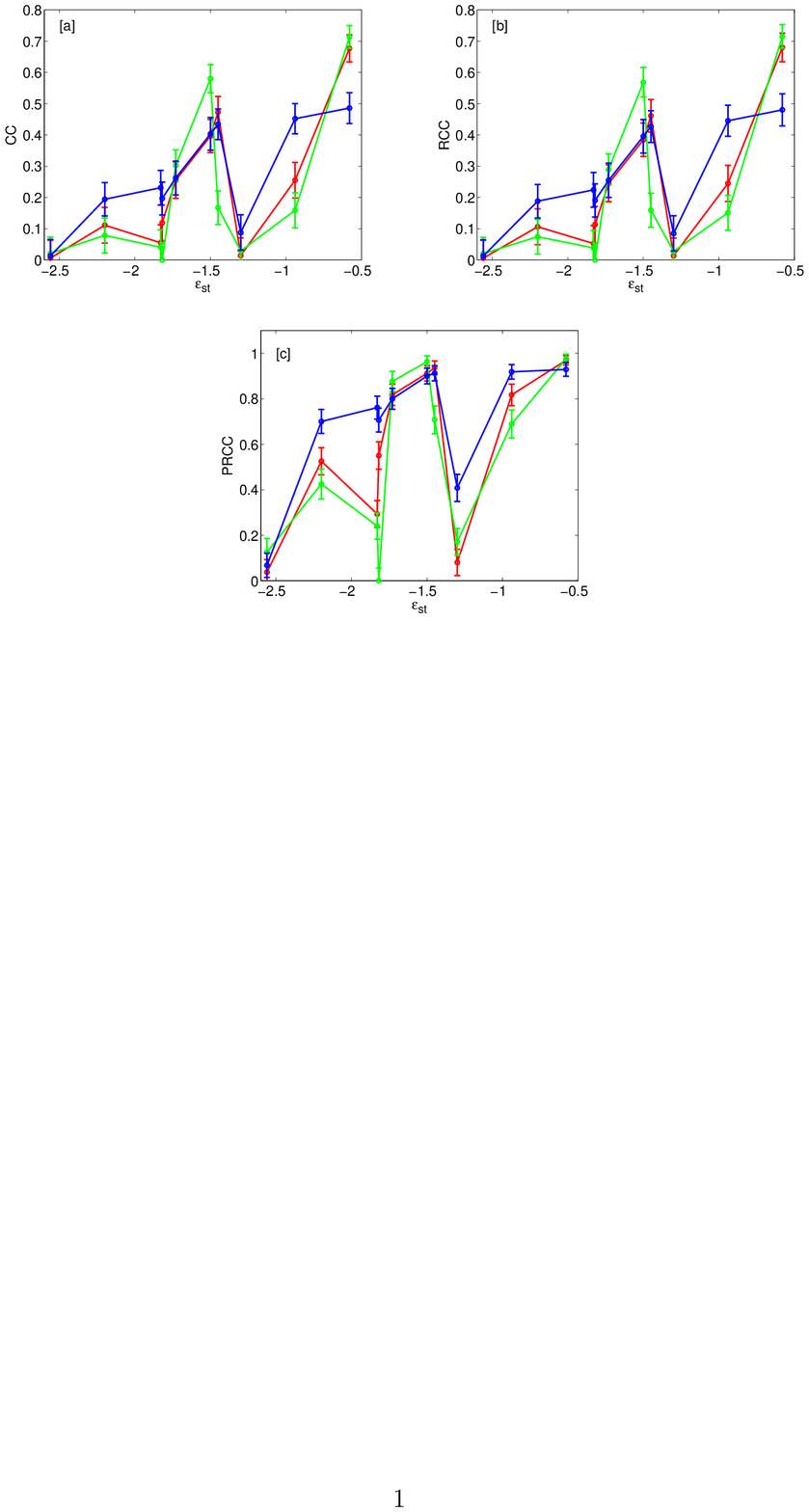}
\caption{(Color online) Correlation coefficients: (a) CC, (b) RCC, and (c) PRCC
versus the free energy of stacking interactions. The red, green, and blue lines
are for the promoter sequence of the T7 phage, the L42B12, and the AdMLP sequences,
respectively. The stacking interactions show similar trends for all the three DNA
sequences.}
\label{fig2}
\end{figure}

As the expression for calculating $ \langle m \rangle $ (see Eq.~(\ref{mb})) is 
monotonic with respect to all the fourteen model parameters, we use the measured
correlation coefficient using different techniques as the measure of sensitivity 
index of these model parameters. 
To this end, we estimate the Pearson correlation coefficient (CC), the Spearman 
or rank correlation coefficient (RCC), and the partial rank correlation coefficient 
(PRCC) (for detail, see Appendix A) for the T7 phage DNA. The 
results are shown in Table~\ref{T7tab}. 
The set of input parameters is chosen in a similar manner as used 
for generating the scatter plots, while calculating correlation coefficients (see 
Appendix A for discussion of the implementation).
Specifically, the correlation coefficients are computed from a set of 100,000 
data points for the generation of input and output data. 
The first set of CC, RCC and PRCC in Table~\ref{T7tab}, designated as 
``All parameters'', are calculated by varying all the fourteen model parameters 
simultaneously.  The values of coefficients obtained from the raw data (CC) 
and rank transformed data (RCC) are very close, as the mean bubble 
size shows a linear dependence on the input parameters. However, the PRCC 
is much higher in magnitude than the CC and the RCC, which signifies that 
the effect of a particular input on the output is not independent 
of the other input parameters. Hence, PRCC gives a clearer picture of the 
parameter sensitivity. 
The set of correlation coefficients for all the stability factors (only 
$\epsilon_{hb}$ and $\epsilon_{st}$)  and for the stacking interactions 
(only $\epsilon_{st}$) are also calculated and are presented in 
Table~\ref{T7tab} under the column heading of ``Excluding $ c $ $ \& $ $ \xi $'' 
and ``Excluding $ c $, $ \xi $ $ \& $ $ \epsilon_{hb} $'', respectively.
These are calculated by perturbing all the parameters except the 
excluded parameters which are kept fixed at the corresponding literature 
values. This is done to check how the parameter sensitivity order changes 
with the gradual decrease in the set of parameter variation. Results show that an
increase in the values of different correlation coefficients occurs (as calculated by 
adopting CC, RCC and PRCC), but the relative order of the sensitivity remains 
the same, which is a signature of the linear dependence of the input parameters. 

\begin{widetext}

\begin{table}
\caption{CC, RCC, and PRCC values for the promoter sequence of the T7 phage
(the importance of the numbers in bold are discussed in Sec III)}
\label{T7tab}
\begin{ruledtabular}
\begin{tabular}{c|ccc|ccc|ccc}
Parameter & & All parameters & & & Excluding        & & & Excluding                                  & \\
                    & &                            & & &  $c$ \& $\xi$    & & & $c$, $\xi$ \& $\epsilon_{hb}$ & \\
 & CC & RCC & PRCC   & CC & RCC & PRCC   & CC & RCC & PRCC  \\
\hline
$\epsilon_{st}$ ($\mathtt{AT-AT}$) & 0.135 & 0.128 & 0.589 & 0.168 & 0.160 & 0.711 & 0.249 & 0.239 & 0.818 \\
$\epsilon_{st}$ ($\mathtt{TA-TA}$) & \textbf{0.358} & \textbf{0.346} & \textbf{0.890} & \textbf{0.436} & \textbf{0.420} & \textbf{0.938} & \textbf{0.676} & \textbf{0.679} & \textbf{0.969} \\
$\epsilon_{st}$ ($\mathtt{AA-TT}$) & 0.207 & 0.199 & 0.752 & 0.257 & 0.245 & 0.843 & 0.395 & 0.382 & 0.913  \\
$\epsilon_{st}$($\mathtt{GA-TC}$) & 0.063 & 0.062 & 0.328 & 0.075 & 0.071 & 0.422 & 0.124 & 0.118 & 0.550  \\ 
$\epsilon_{st}$($\mathtt{CA-TG}$) & 0.134 & 0.129 & 0.594 & 0.168 & 0.160 & 0.709 & 0.256 & 0.246 & 0.819  \\ 
$\epsilon_{st}$($\mathtt{AG-CT}$) & 0.251 & 0.242 & 0.807 & 0.310 & 0.297 & 0.882 & 0.471 & 0.459 & 0.937  \\ 
$\epsilon_{st}$($\mathtt{AC-GT}$) & 0.054 & 0.051 & 0.301 & 0.073 & 0.070 & 0.403 & 0.112 & 0.108 & 0.529  \\ 
$\epsilon_{st}$($\mathtt{GG-CC}$) & 0.030 & 0.030 & 0.158 & 0.035 & 0.032 & 0.209 & 0.052 & 0.050 & 0.295 \\ 
$\epsilon_{st}$($\mathtt{CG-CG}$) & 0.012 & 0.011 & 0.038 & 0.010 & 0.010 & 0.063 & 0.014 & 0.013 & 0.084  \\ 
$\epsilon_{st}(\mathtt{GC-GC})$ & 0.002 & 0.002 & 0.022 & 0.000 & 0.001 & 0.031 & 0.009 & 0.008 & 0.034  \\ 
$\epsilon_{hb}(\mathtt{AT})$ & \textbf{0.624} & \textbf{0.622} & \textbf{0.961} & \textbf{0.760} & \textbf{0.768} & \textbf{0.980} &  &  &   \\ 
$\epsilon_{hb}(\mathtt{GC})$ & 0.079 & 0.076 & 0.390 & 0.096 & 0.091 & 0.502 &  &  &   \\ 
$\xi$ & \textbf{0.556} & \textbf{0.550} & \textbf{0.951} &  &  &  &  &  &   \\ 
c & -0.101 & -0.096 & -0.467 &  &  &  &  &  &   \\ 
\end{tabular}
\end{ruledtabular}
\end{table}

\end{widetext}

It is further evident from Fig.~\ref{fig1} and Table~\ref{T7tab} that the hydrogen 
bond interaction energy for an $\mathtt{AT}$ base pair $\epsilon_{hb}(\mathtt{AT})$
and the ring 
factor $\xi$ are highly sensitive among the set of model parameters controlling
the breathing dynamics.
However, it is also interesting to analyze the sensitivity of different stacking 
interactions on the breathing dynamics.
The parameter $\epsilon_{st}(\mathtt{TA-TA})$ shows the highest degree of correlation among all the stacking 
energies, which is actually the weakest interaction in DNA double helix. Frequent 
bubble events in the weaker $\mathtt{TATA}$ motif, a key element in this T7
promoter sequence, also justify the SENSA due to
the effected greater probability of bubble formation. 
The order of sensitivity of the parameters for stacking interaction, as found out by the calculation 
of correlation coefficients, for the sequence T7 phage promoter is thus as follows
\begin{eqnarray}
\label{T7sen}
&\epsilon_{st}(\mathtt{TA-TA})>\epsilon_{st}(\mathtt{AG-CT})> \epsilon_{st}(\mathtt{AA-TT})>\epsilon_{st}(\mathtt{AT-AT}) \nonumber \\
& \approx \epsilon_{st}(\mathtt{CA-TG})>\epsilon_{st}(\mathtt{GA-TC})>\epsilon_{st}(\mathtt{AC-GT})> 
\epsilon_{st}(\mathtt{GG-CC}) \nonumber \\
& > \epsilon_{st}(\mathtt{CG-CG})>\epsilon_{st}(\mathtt{GC-GC}) .
\end{eqnarray}

\noindent
Eq.~(\ref{T7sen}) shows that sensitivity of the stacking interaction energies involved 
with AT base pair are higher compared to the stacking interaction energies between the two 
neighboring GC bases. 
One of the reasons may be the presence of fewer numbers of $\epsilon_{st}(\mathtt{
CG-CG})$ and $\epsilon_{st}(\mathtt{GC-GC})$ in the promoter sequence of T7 phage
(both appear only twice). However, another stacking interaction between two
neighboring $\mathtt{GC}$ ($\epsilon_{st}(\mathtt{GG-CC})$) appears relatively
frequently in the sequence. Thus one 
cannot generalize the effectivity of a particular stacking interaction energy on the breathing 
dynamics from the point of view of the number of appearance of that particular 
stacking interaction, and thus we are led to conclude that both the number of
occurrences and the nature of the stacking interaction affect the sensitivity order.
To check how the relative sensitivity order of these model breathing parameters, mainly the 
stacking interactions, gets altered due to the sequence of hetero-polymeric DNA, we performed the 
above mentioned calculation for the correlation coefficients (CC, RCC, PRCC) on two 
other different DNA sequences. These two other DNA hetero-polymeric 
chains are L42B18 and AdMLP with the following sequences
\begin{widetext}
\begin{eqnarray}
\label{l42}
\texttt{5'-cCGCCAGCGGCGTTAATACTTAAGTATTATGGCCGCTGCGCc -'3} ,
\end{eqnarray}
and
\begin{eqnarray}
\label{AdMLP}
\texttt{5'-gCCACGTGACCAGGGGTCCCCGCCGGGGGGGTATAAAAGGGGCGGACC}
\nonumber \\
\texttt{TCTGTTCGTCCTCACTGTCTTCCGGATCGCTGTCCAg -'3} .
\end{eqnarray}
\end{widetext}

The list of correlators: CC, RCC, and PRCC obtained by varying all the model
parameters, for the two sequences (\ref{l42}) and (\ref{AdMLP}) are listed in 
Tab.~\ref{l42tab} and Tab.~\ref{Adtab}. The correlation coefficients related
to $\xi$ and $c$ should be independent of the sequence as indeed observed. In
the case of the hydrogen bond energies, $\epsilon_{hb}(\mathtt{AT})$ has a much
higher correlation coefficient compared to $\epsilon_{hb}(\mathtt{GC})$ for all
three sequences.  However, in AdMLP the value of CC (also RCC and PRCC) for
$\epsilon_{hb}(\mathtt{GC})$ is higher than the other two sequences, as in AdMLP
the $\mathtt{GC}:\mathtt{AT}$ number ratio is much higher than in the other two
sequences. The stacking interactions in hierarchical order of sensitivity for
the L42B12 and AdMLP DNA sequences are
\begin{eqnarray}
\label{l42b18sen}
\epsilon_{st}(\mathtt{TA-TA})>\epsilon_{st}(\mathtt{AA-TT})> 
\epsilon_{st}(\mathtt{AT-AT})> \epsilon_{st}(\mathtt{AG-CT}) \nonumber \\
> \epsilon_{st}(\mathtt{CA-TG})>
\epsilon_{st}(\mathtt{AC-GT})> \epsilon_{st}(\mathtt{GG-CC}) \nonumber \\
> \epsilon_{st}(\mathtt{CG-CG})>\epsilon_{st}(\mathtt{GC-GC})>
\epsilon_{st}(\mathtt{GA-TC}),
\nonumber \\
\end{eqnarray}

\noindent and
\begin{eqnarray}
\label{AdMLPsen}
\epsilon_{st}(\mathtt{TA-TA})>\epsilon_{st}(\mathtt{CA-TG})\sim \epsilon_{st}(\mathtt{AG-CT})>\epsilon_{st}(\mathtt{AA-TT}) \nonumber \\
> \epsilon_{st}(\mathtt{AT-AT})>
\epsilon_{st}(\mathtt{GG-CC})>\epsilon_{st}(\mathtt{GA-TC}) \nonumber \\
> \epsilon_{st}(\mathtt{AC-GT})> 
\epsilon_{st}(\mathtt{CG-CG})>\epsilon_{st}(\mathtt{GC-GC}) .
\nonumber \\
\end{eqnarray}

\noindent By comparing the relative sensitivity order of the three analyzed sequences,
the sensitivity of stacking interaction between two neighboring GC
bases ($\epsilon_{st}(\mathtt{GG-CC})$, $\epsilon_{st}(\mathtt{CG-CG})$,
and $\epsilon_{st}(\mathtt{GC-GC})$) are found to be very low for T7
and L42B18, but in AdMLP the correlation coefficient value (see in Table
\ref{Adtab}) of $\epsilon_{st}(\mathtt{GG-CC})$  is relatively high. This
happens as the AdMLP sequence has higher $\mathtt{GC}$ content and the
$\epsilon_{st}(\mathtt{GG-CC})$ stacking interaction appears 24 times in
this sequence. But $\epsilon_{st}(\mathtt{TA-TA})$ still shows the highest
correlation value among all stacking interaction energies for all three
sequences though it appears only twice in the AdMLP sequence. This is a
signature of the fact that the $\mathtt{TA-TA}$ stacking interaction has
an overriding influence on the breathing process even in a situation where
its numbers are low. A table containing the number of appearance of each
stacking interaction in all the three sequences is given for reference (see
Table ~\ref{numtab}).  Our objective to perform the SENSA for the sequences
L42B12 and AdMLP along with the T7 phage promoter sequence is to figure out
how the sensitivity order of the stacking interaction parameters changes with
the variation of DNA sequences. For this purpose a pictorial presentation of
the free energy of stacking interactions versus their correlation coefficients
(CC, RCC and PRCC) are given for all three DNA sequences in Fig.\ref{fig2}:
we see that the general trend of the sensitivity order remains more or
less unchanged in these three different DNA sequences. One may thus group
out the stacking interactions such that $\epsilon_{st}(\mathtt{AT-AT})$,
$\epsilon_{st}(\mathtt{TA-TA})$, $\epsilon_{st}(\mathtt{AA-TT})$,
$\epsilon_{st}(\mathtt{CA-TG})$, and $\epsilon_{st}(\mathtt{AG-CT})$
are more sensitive towards bubble opening than the other five stacking
interactions. This results together lead us to conclude that the nature of
the stacking interaction is predominant over the number of appearance of
that particular stacking interaction in calculating the sensitivity for a
hetero-polymeric DNA sequence.

\section{Discussion and Conclusion}

Sensitivity analysis can help in a rigorous study of a system. We
quantified the sensitivity of bubble formation with respect to the stability
parameters (as given by the Poland-Scheraga model), which leads us to a better
understanding of the sequence dependent nature of the breathing dynamics of
hetero-polymeric DNA. The general trend of this parameter SENSA as
evaluated from our calculations is that it does not significantly depend
on the nature of the DNA sequence. We showed that the
number of occurrences of a particular interaction (hydrogen bond interaction
or stacking interaction) is not the major factor in the degree of sensitivity.
Rather, the specific nature of a particular interaction is the
major player, even in a situation where its number of occurrences
in a DNA sequence is smaller. Generally, the SENSA also shows that
the bubble opening free energy $\xi$ and the hydrogen bonding free energy
$\epsilon_{hb}(\mathtt{AT})$ are always highly sensitive parameters.
These results will help in a better
understanding of the relative probability of bubble opening and how it varies
with the change in DNA sequences.

The sensitivity data, as revealed and discussed in the previous section, has
its own role in grading the different interaction types in order of importance,
but the information can be used for other important studies as well, like
an optimization problem to find out the correct values of the breathing
dynamics parameters.  The SENSA data if used properly can have a significant
influence during parameter optimization of the system.  All parameters may
not equally affect or influence the output. If one exploits the parameters
having higher sensitivity more than the other parameters during optimization,
the convergence may occur faster than a simulation in which all parameters
are searched with equal weights.

\begin{widetext}

\begin{table}
\caption{CC, RCC, and PRCC values of the L42B18 sequence (the importance of the
numbers in bold are discussed in Sec III)}
\label{l42tab}
\begin{ruledtabular}
\begin{tabular}{c|ccc|ccc|ccc}
Parameter & & All parameters & & & Excluding        & & & Excluding                                  & \\
                    & &                            & & &  $c$ \& $\xi$    & & & $c$, $\xi$ \& $\epsilon_{hb}$ & \\
 & CC & RCC & PRCC   & CC & RCC & PRCC   & CC & RCC & PRCC  \\
\hline
 $\epsilon_{st}(\mathtt{AT-AT})$ & 0.170 & 0.163 & 0.681 & 0.202 & 0.192 & 0.772 & 0.300 & 0.287 & 0.876  \\ 
 $\epsilon_{st}(\mathtt{TA-TA})$ & \textbf{0.404} & \textbf{0.393} & \textbf{0.912} & \textbf{0.476} & \textbf{0.461} & \textbf{0.946} & \textbf{0.715} & \textbf{0.716} &  \textbf{0.976}  \\ 
 $\epsilon_{st}(\mathtt{AA-TT})$ & 0.333 & 0.323 & 0.873 & 0.389 & 0.374 & 0.920 & 0.578 & 0.566 &  0.963 \\ 
 $\epsilon_{st}(\mathtt{GA-TC})$ & 0.006 & 0.006 & 0.003 & 0.005 & 0.006 & 0.004 & 0.002 & 0.002 & 0.000 \\ 
 $\epsilon_{st}(\mathtt{CA-TG})$ & 0.093 & 0.087 & 0.441 & 0.105 & 0.100 & 0.536 & 0.158 & 0.151 & 0.689 \\ 
 $\epsilon_{st}(\mathtt{AG-CT})$ & 0.096 & 0.091 & 0.456 & 0.114 & 0.108 & 0.555 & 0.167 & 0.158 & 0.705  \\ 
 $\epsilon_{st}(\mathtt{AC-GT})$ & 0.040 & 0.039 & 0.237 & 0.052 & 0.050 & 0.303 & 0.074 & 0.069 & 0.421 \\ 
 $\epsilon_{st}(\mathtt{GG-CC})$ & 0.023 & 0.021 & 0.122 & 0.027 & 0.025 & 0.160 & 0.033 & 0.032 & 0.243 \\ 
 $\epsilon_{st}(\mathtt{CG-CG})$ & 0.016 & 0.015 & 0.089 & 0.018 & 0.017 & 0.114 & 0.024 & 0.023 &  0.171 \\ 
 $\epsilon_{st}(\mathtt{GC-GC})$ & 0.008 & 0.008 & 0.065 & 0.012 & 0.011 & 0.085 & 0.025 & 0.024 & 0.126  \\ 
 $\epsilon_{hb}(\mathtt{AT})$ & \textbf{0.633} & \textbf{0.633} & \textbf{0.963} & \textbf{0.742} & \textbf{0.748} & \textbf{0.978} &  &  &   \\ 
 $\epsilon_{hb}(\mathtt{GC})$ & 0.039 & 0.037 & 0.241 & 0.056 & 0.053 & 0.303 &  &  &   \\ 
 $\xi$  & 0.508 & 0.500 & 0.943 &  &  &  &  &  &   \\ 
 c  & -0.108 & -0.104 & -0.514 &  &  &  &  &  &   \\ 
\end{tabular}
\end{ruledtabular}
\end{table}

\begin{table}
\caption{CC, RCC, and PRCC values of the AdMLP sequence (the importance of the
numbers in bold are discussed in Sec III)}
\label{Adtab}
\begin{ruledtabular}
\begin{tabular}{c|ccc|ccc|ccc}
Parameter & & All parameters & & & Excluding        & & & Excluding                                  & \\
                    & &                            & & &  $c$ \& $\xi$    & & & $c$, $\xi$ \& $\epsilon_{hb}$ & \\
 & CC & RCC & PRCC   & CC & RCC & PRCC   & CC & RCC & PRCC  \\
\hline
 $\epsilon_{st}(\mathtt{AT-AT})$ & 0.128 & 0.123 & 0.575 & 0.160 & 0.152 & 0.699 & 0.265 & 0.258 & 0.797  \\ 
 $\epsilon_{st}(\mathtt{TA-TA)}$ & \textbf{0.242} & \textbf{0.233} & \textbf{0.792} & \textbf{0.306} & \textbf{0.294} & \textbf{0.877} & \textbf{0.482} & \textbf{0.476} &	\textbf{0.927} \\ 
 $\epsilon_{st}(\mathtt{AA-TT})$ & 0.196 & 0.187 & 0.730 & 0.253 & 0.241 & 0.833 & 0.405 & 0.397 & 0.898 \\ 
 $\epsilon_{st}(\mathtt{GA-TC})$ & 0.096 & 0.092 & 0.466 & 0.122 & 0.115 & 0.589 & 0.198 & 0.191 & 0.702 \\ 
 $\epsilon_{st}(\mathtt{CA-TG})$ & 0.219 & 0.210 & 0.768 & 0.281 & 0.270 & 0.861 & 0.453 & 0.446 & 0.918 \\ 
 $\epsilon_{st}(\mathtt{AG-CT})$ & 0.216 & 0.208 & 0.756 & 0.273 & 0.261 & 0.850 & 0.437 & 0.430 & 0.911 \\ 
 $\epsilon_{st}(\mathtt{AC-GT})$ & 0.096 & 0.091 & 0.455 & 0.122 & 0.116 & 0.584 & 0.195 & 0.188 & 0.695 \\ 
 $\epsilon_{st}(\mathtt{GG-CC})$ & \textbf{0.113} & \textbf{0.108} & \textbf{0.523} & \textbf{0.146} & \textbf{0.139} & \textbf{0.651} & \textbf{0.235} & \textbf{0.228} & \textbf{0.759} \\ 
 $\epsilon_{st}(\mathtt{CG-CG})$ & 0.042 & 0.040 & 0.225 & 0.051 & 0.048 & 0.313 & 0.083 & 0.080 & 0.404 \\ 
 $\epsilon_{st}(\mathtt{GC-GC})$ & 0.027 & 0.023 & 0.034 & 0.065 & 0.056 & 0.047 & 0.015 & 0.015  & 0.069 \\ 
 $\epsilon_{hb}(\mathtt{AT})$ & \textbf{0.587} & \textbf{0.583} & \textbf{0.956} & \textbf{0.748} & \textbf{0.756} & \textbf{0.978} &  &  & \\ 
 $\epsilon_{hb}(\mathtt{GC})$ & \textbf{0.176} & \textbf{0.169} & \textbf{0.674} & \textbf{0.211} & \textbf{0.201} & \textbf{0.787} &  &  &  \\ 
 $\xi$  & 0.612  & 0.609 & 0.960 &  &  &  &  &  & \\ 
 c  & -0.098 & -0.093 & -0.465 &  &  &  &  &  & \\ 
\end{tabular}
\end{ruledtabular}
\end{table}

\end{widetext}

\begin{figure}[!t]
\includegraphics[width=0.75\columnwidth,angle=0,clip=]{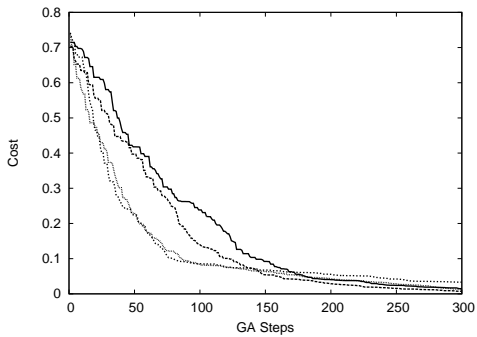}
\caption{Cost profile obtained for different cases of optimization. The solid
line represents unconstrained optimization, the dashed and dotted lines are
obtained by keeping the cumulative probability for mutation of the sensitive
parameters at 20$ \% $ and 60$ \% $, respectively. The close-dotted line is
generated by keeping this cumulative probability at 60 $ \% $ up to 200 GA
steps, and then at 20$ \% $ for the rest of the optimization.}
\label{fig-cost}
\end{figure}

We optimized all parameters
associated with the breathing dynamics taking the equilibrium distribution
function generated by mimicking the experimental scenario \cite{altan}, as the
objective during optimization (see Appendix B). We used the Genetic Algorithm
(GA) \cite{gold} as the optimizer and the promoter sequence of T7 Phage for
this study. Fig.\ref{fig-cost} represents the cost profile versus the number of GA
steps for different runs with different extent of constraints in searching the
sensitive parameters. The cost function (see Appendix B) is a measure of how
close we are to obtaining our solution. When the cost tends to zero the actual
solution is found out. The solid line is the profile for the normal optimization,
during which no additional condition is imposed on the optimizer. But the
other lines represent cost profiles, for which the optimization was biased to
sample the sensitive parameters more by allowing it to mutate or get sampled
more than the others. In GA, mutation occurs with a probability (mutation
probability) which is set initially. Generally there is no bias in the choice
of variables for mutation. We incorporated a condition in the choice
of variable selected to undergo mutation. The more sensitive parameters
have a higher probability to be selected for mutation. In Fig.\ref{fig-cost}
the dashed line represents the profile where, the cumulative probability for
$\xi$ and $\epsilon_{hb}(\mathtt{AT})$ (the most sensitive parameters) to be chosen
for mutation is 20$\%$ and the rest is the cumulative probability for the
other parameters. The cost profile with 60$\%$ of such a cumulative probability
for the sensitive parameters is denoted by the dotted line. The dashed line
(20$\%$) falls at a greater rate than the solid one, but the gain in the initial
convergence is much more prominent for the dotted line (60$\%$). In spite of
this gain, the plateau region in the optimization profile starts at a higher
cost in the case of the dotted line than for the solid line. The probable reason may be
the incomplete search of the other relatively less sensitive parameters. Thus
we designed the optimization scheme such that initially the search would
be highly biased towards the sensitive parameters and after certain steps
of optimization, the cumulative probability to be picked up for mutation, of
those higher sensitive parameters would be decreased. The close-dotted line in
Fig.\ref{fig-cost} was generated by keeping this cumulative probability
60$ \% $ initially (upto 200 GA steps) and then it is decreased to 20$ \% $.
This strategy of gradual reduction in sampling importance (kept high initially
and decreased later) of the more sensitive parameters is the ideal strategy
to handle the present problem in a more computationally cost-efficient way.


\begin{acknowledgments}
ST and PC acknowledge financial support from UGC, New Delhi, for granting 
Junior Research Fellowship.  
SS thanks UGC, New Delhi, for granting D. S. Kothari post-doctoral fellowship. 
RM acknowledges funding through the Academy of Finland's FiDiPro scheme.
\end{acknowledgments}

\begin{table}
\caption{List of the number of occurrance of different stacking interactions in
the three studied DNA sequences (the importance of the numbers in bold are
discussed in Sec III)}
\label{numtab}
\begin{ruledtabular}
\begin{tabular}{ccccccc}
Stacking interaction &$ \epsilon_{st} $ & T7 & & L42B18 & & AdMLP \\
\hline  $(\mathtt{AT-AT})$& 1.729409 & 4 & & 3 & & 2 \\ 
 $(\mathtt{TA-TA})$ & 0.579800 & 6 & & 5 & & \textbf{2} \\ 
  $(\mathtt{AA-TT})$ & 1.499484 & 7 & & 5 & & 5 \\ 
 $(\mathtt{GA-TC})$ & 1.819371 & 11 & & 0 & & 12 \\ 
 $(\mathtt{CA-TG})$ & 0.939677 & 7 & & 3 & & 8 \\ 
 $(\mathtt{AG-CT})$ & 1.455363 & 14 & & 4 & & 9 \\ 
 $(\mathtt{AC-GT})$ & 2.199241 & 9 & & 3 & & 11 \\ 
 $(\mathtt{GG-CC})$ & 1.829370 & 7 & & 6 & & \textbf{24} \\ 
 $(\mathtt{CG-CG})$ & 1.299554 & 2 & & 5 & & 7 \\ 
 $(\mathtt{GC-GC})$ & 2.559130 & 2 & & 7 & & 4 \\ 
\end{tabular} 
\end{ruledtabular}
\end{table}

\appendix

\section{Correlation Coefficient}

A simple way to examine a given parameter's sensitivity is to obtain the degree
of correlation of various input parameters to the output.
The Correlation Coefficient is a quantity to measure how strong the output of a system is linearly associated with the particular input parameter. It also accounts for the direction (positive or negative) of this linear association.
Thus it may be used as sensitivity index for a system in which the output varies linearly with the input variables, as it actually accounts for the perturbation on the output when input parameters are varied. The
Correlation Coefficient of a particular parameter is fundamentally the measure of covariance between the output and that of input parameter, which is then normalised by dividing with the product of the standard deviation of input and output,
\begin{equation}
\label{cc}
r_{x_{j}y} =
\frac{
\sum_{i=1}^{N} (x_{ij}-\bar{x}_j)(y_{i}-\bar{y})
}{
\sqrt{\sum_{i=1}^{N} (x_{ij}-\bar{x}_j)^2 \sum_{i=1}^{N}(y_{i}-\bar{y})^2}
}.
\end{equation}
Here $r_{x_{j}y}$ is the correlation coefficient of the input 
parameter $x_j$ and output $y$, $ \bar{x}_j $ and $ \bar{y} $ are the mean of input and output, respectively and $ N $ is the number of sampling. The value of $r_{x_{j}y}$ varies from -1 to +1. 
The `+' or `-' sign denotes the direction of the linear dependence, i.e, whether the output data increases or decreases with the increase in input parameter.
For higher magnitudes of 
$r_{x_{j}y}$ the effect of that particular input parameter will be larger on the output parameter. 
Then that particular input parameter is said to be highly sensitive with respect to that output. 
A very low value of the correlation coefficient means that the output will differ only a little 
even when the perturbation on input is very high which signifies the lesser sensitivity of that input. If $r_{x_{j}y}$ is calculated from the raw 
data of input and output using Eq. (\ref{cc}), it is known as Pearson correlation coefficient (CC).

The parameter dependence may not always be linear. The Pearson correlation coefficient is 
inadequate to show the actual picture of sensitivity of the system in that case. If one uses 
rank transformed data instead of the raw data of both the input and output parameters to 
calculate the correlation coefficient (known as rank correlation coefficient (RCC)), it will 
account for the nonlinear, yet monotonic trend of parameter dependence. The formal name 
of RCC is Spearman Correlation Coefficient. 

The
Partial Rank correlation coefficient (PRCC) accounts for the dependence of a particular input 
with the output after deducting the effect of other inputs. PRCC of a set of inputs $x_j$ 
and output $y$ may be calculated as the RCC of $x_j-\tilde{x}_j$ and $y-\tilde{y}$, where 
$\tilde{x}_j$ and $\tilde{y}$ account for the effect of other input parameters on that particular input $ x_j $ and output $ y $. These can be measured following the regression model \cite{marino}
\begin{equation}
\tilde{x}_j=c_0+\sum_{l\neq{j}}^k c_lx_l, \;
\tilde{y}=b_0+\sum_{l\neq{j}}^k b_lx_l .
\end{equation}
The PRCC can also be expressed in terms of the rank correlation matrix 
($\mathcal{C}$). The matrix element $\mathcal{C}_{ij}$ represents the RCC between 
the $i$th and $j$th components. If $P_{ij} $ is the co-factor of $\mathcal{C}_{ij}$, then 
PRCC ($\mathcal{P}_{ij}$) will be \cite{cramer}
\begin{equation}
\mathcal{P}_{ij}=-\frac{P_{ij}}{\sqrt{P_{ii}P_{jj}}} .
\end{equation}
Thus the PRCC of input parameters $x_j$ with respect to some output parameter
$y$ of a system can be written as
\begin{equation}
\label{prcc}
\mathcal{P}_{x_jy}=-\frac{P_{x_jy}}{\sqrt{P_{x_jx_j}P_{yy}}} .
\end{equation}
The correlation coefficient may give the picture of 
sensitivity properly only if the change of output with the input is monotonic. For non-monotonic 
relation one may perform variance based sensitive test.

The implementation of the the correlation coefficient estimation is as follows.
We generate the set of data points of input parameters by randomly perturbing it within the range of $ \pm $5$ \% $ of the reported literature value \cite{krueger,richard}. The output data is calculated using the perturbed input variables. To estimate the CC, these set of input and output data are put into the expression (\ref{cc}). For the RCC calculation both the input and output data are arranged in an increasing or decreasing order and a rank is set for each data. Then the correlation coefficient is calculated with that of rank transformed data.
For PRCC calculation we used the second procedure with Eq.~(\ref{prcc}) among the two above mentioned techniques. The
RCC between the different input parameters as well as the RCC between the inputs and the output are calculated.  These RCC values could then be arranged in matrix $\mathcal{C}$. The PRCC is calculated with the co-factors of this matrix by using Eq.~(\ref{prcc}).

\section{Genetic Algorithm}

Mimicking the experimental scenario of Ref.~\onlinecite{altan},
one arrives at the equilibrium probability distribution for fluorophore 
tagged base pairs ($ x_T$) using Eq.~(\ref{eqd}). 
Fluorescence signals appear if the base pairs in the $ \delta $ neighbourhood of the fluorophore are open. The time dynamics of the occurrence of fluorescence is thus related to the breathing dynamics, as local denaturation of all base pairs in $ x_T \pm \delta $ is necessary for appearing fluoresce signal (in our calculation we take 
$ \delta =0$).

To obtain the DNA stability parameters of DNA by optimization, the objective function may be defined as \cite{talukder1}
\begin{equation}
cost=\sum_{i=1}^M (P_{e}(x_i)-P(x_i))^2 ,
\end{equation}

\noindent
where $ P_{e}(x_i) $ is the equilibrium probability of the tagged base pair at the $i${th} position in the sequence for the experimental value of the input parameters and likewise $ P(x_i) $ is the equilibrium probability for the set of input parameters obtained in a step, during optimization. The $ cost $ is actually the difference in these probabilities ($ P_{e}(x_i)-P(x_i) $) and in the course of optimization it decreases. We reach our solution when $ cost\rightarrow0 $. 

We apply the Genetic Algorithm (GA) to optimize the parameters involved in the
breathing dynamics. In GA the cost function is replaced by the fitness function,
\begin{equation}
f = \exp(-cost),
\end{equation}
such that a decrease in cost leads to an increase in the fitness function. At the
end of the simulation $f$ approaches 1. The progress towards achieving $f
\rightarrow1$ in the GA occurs by repeated use of three operations, namely,
selection, crossover and mutation. These
operators closely mimic similar biological processes in conventional genetics.
Since GA mimics these natural processes, it is sometimes referred to as a natural
algorithm for optimization.

\end{document}